\documentclass[twocolumn]{aastex631}

\usepackage{graphicx} 
\usepackage{lineno}

\usepackage{graphicx} 
\usepackage{natbib}

\submitjournal{ApJ}

\begin{document}


\title{Determining Total Infrared Luminosities from Submm Measurements of High Redshift Galaxies }

\correspondingauthor{Mar\'ia Emilia De Rossi}
\email{mariaemilia.dr@gmail.com}

\author[0000-0002-4575-6886]{Mar\'ia Emilia De Rossi}
\affiliation{ Universidad de Buenos Aires, Facultad de Ciencias Exactas y Naturales y Ciclo Básico Común, Buenos Aires, Argentina}

\affiliation{  CONICET-Universidad de Buenos Aires, Instituto de Astronomía y Física del Espacio (IAFE), Buenos Aires, Argentina}

\author[0000-0003-2303-6519]{George H. Rieke}
\affiliation{Steward Observatory, University of Arizona,
933 North Cherry Avenue, Tucson, AZ 85721, USA}

\author[0000-0003-2919-7495]{Christina C. Williams}
\affiliation{NSF National Optical-Infrared Astronomy Research Laboratory, 950 N. Cherry Ave., Tucson, AZ 85719, USA}

\author[0000-0003-0212-2979]{Volker Bromm}
\affiliation{Department of Astronomy,University of Texasat Austin, 2515Speedway, Stop C1400,Austin,TX 78712,USA}

\affiliation{Cosmic Frontier Center, The University of Texas at Austin, Austin,TX 78712,USA}

\affiliation{Weinberg Institute for Theoretical Physics, University of Texas at Austin, Austin,TX 78712,USA}

\author[0000-0002-6221-1829]{Jianwei Lyu}
\affiliation{Steward Observatory, University of Arizona, 933 North Cherry Avenue, Tucson, AZ 85721, USA}

\author[0000-0002-9720-3255]{Meredith A. Stone}
\affiliation{Steward Observatory, University of Arizona, 933 North Cherry Avenue, Tucson, AZ 85721, USA}

\author[0000-0001-6561-9443]{Yang Sun}
\affiliation{Steward Observatory, University of Arizona,
933 North Cherry Avenue, Tucson, AZ 85721, USA}

\author[0000-0001-9262-9997]{Christopher N.\ A.\ Willmer}
\affiliation{Steward Observatory, University of Arizona, 933 North Cherry Avenue, Tucson, AZ 85721, USA}

\author[0000-0003-3307-7525]{Yongda Zhu}
\affiliation{Steward Observatory, University of Arizona, 933 North Cherry Avenue, Tucson, AZ 85721, USA}

\begin{abstract}

Determining total infrared luminosities for very high redshift galaxies is important to estimate the rate of star formation in heavily dust-embedded environments. It is also challenging because the most sensitive far infrared observatory, Herschel, was limited in sensitivity and its deepest measurements are subject to confusion noise. Thus, these determinations largely depend on ALMA, observing in the mm- and/or submm-wavelengths, which sample only the long-wavelength part of the spectral energy distributions (SEDs). Luminosities are conventionally estimated with modified blackbody fits to these measurements,  but do not include the emission in the mid-infrared by warmer dust;  there is evidence that this mid-IR component may be relatively strong in very high-redshift galaxies compared with local ones.  A correction factor must be applied to the modified black body luminosities to derive the total infrared luminosity. We study infrared SEDs using simulations tuned to galactic conditions typical of high-z galaxies.  We find that the different behaviors of infrared SEDs are dominated by a single  key physical parameter, the luminosity density. This allows us to estimate the corrections for the missing mid-infrared luminosity in a general way. We find that a factor of $\sim$ 1.6 - 1.7 (0.2 dex) is appropriate in most circumstances, with a larger factor of $\sim$ 1.75 - 1.85 ($\sim$ 0.25 dex) up to 2 (0.3 dex) necessary for  high redshift (z $>$ 4) galaxies at the highest luminosities, $> 10^{12}$ L$_\odot$.  These corrections are needed to estimate star formation rates based on total infrared luminosity.

\end{abstract}


\section{Introduction}

A long-standing goal of extragalactic astronomy is to map the totality of star formation throughout cosmic history \citep[e.g.,][]{madau2014}. In this context, determining the far- and total-infrared luminosities is central to estimating the star formation rates (SFRs) of dusty galaxies \citep[e.g.,][]{kennicutt1998, casey2014}. A full understanding of the far infrared spectral energy distributions (SEDs) is needed to reach this goal. ALMA has provided many measurements on the Rayleigh-Jeans part of the SEDs, but cannot cover the critical shorter far infrared wavelengths.  It therefore has modest  ability to measure SED temperatures but very little to determine SED shapes, e.g., widths. A recent summary of ALMA results is provided by \citet{mitsuhashi2024a}, finding at $z \sim 6 - 7$ a dust temperature  of $T_{\rm dust} = 40.9^{+10.0}_{-9.1}$ K, from which the authors conclude that there is ``a gentle increase of $T_{\rm dust}$ from $z = 0$ to $z \sim 6 - 7$.'' 
For examples of individual galaxies measured in multiple far infrared bands with results supporting the temperature increase, see, e.g.,  \citet[][]{bakx2021, tripodi2023} and for a summary of relevant observations, see \citet{villanueva2024}. Because the majority of galaxies in their sample are measured at only two far infrared bands, \citet{mitsuhashi2024a} could not test for changes in the shape of the SED, e.g. for SED broadening. 

Without spectral coverage to shorter infrared wavelengths, these measurements only loosely constrain the spectral energy distributions (SEDs), which compromises their ability to measure total infrared luminosities. In general, the set of measurements at the shorter far infrared wavelengths is  sparse at high redshift ($z \gtrsim 4$), so one resorts to spectral templates that can be normalized to the available measurements and then integrated to obtain luminosities. A common approach is to fit the data with a blackbody modified to have declining emissivity towards long wavelengths \citep{dunne2000,beelen2006}. 
However, such models leave out the significant fraction of luminosity emitted at wavelengths shorter than the Wien cutoff of the blackbody. Where there is sufficient wavelength coverage, this luminosity can be recovered, e.g., by fitting power law approximations to the  measurements at $\lambda_{\rm rest} \le 70 \,\mu$m \citep{casey2012}. However, these measurements are seldom available at high redshifts making the fits very underconstrained.

In principle, more accurate estimates can be made using spectral templates that include the average behavior of real galaxies. The available templates are almost entirely determined for local or modest redshift galaxies ($z \lesssim 3 $), but it is clear that the SEDs, and hence optimal templates for typical galaxies, should evolve with redshift \citep[e.g.,][]{nordon2010, rujo2013, schreiber2018}.  However, with only ALMA data at $\sim$ 1\,mm (observed), it is difficult to constrain the selection of templates. 

 \citet{derossi2018} analyzed all the cases where sufficient measurements were available to constrain the far infrared SED peak, i.e. at wavelengths shorter than and longer than this feature. The results are summarized in Table~\ref{tab:chisq}. Few additional relevant measurements at the highest redshifts have become available, because of the decommissioning of the Herschel mission. 
\citet{derossi2018} found a temperature increase in the SED by of order 10~K to the range $40 - 50$~K from low redshifts to 5 $<$ z $<$ 7 and a substantial broadening of the SED. Specifically, they found that using the SED of Haro~11, a local low-metallicity, extreme star-forming galaxy \citep[see e.g.,][]{adamo2010,cormier2012, menacho2019}, provided a good fit for the average behavior of the $z \sim 6$ galaxies; its peak corresponds to a temperature of $ \approx $ 44 K, and it is significantly broader in wavelength than templates at similar luminosity derived at low redshift.  We refer readers to \cite{derossi2018} for further information.  This behavior is qualitatively similar to that found by \citet{schreiber2018}, and the temperature shift with increasing redshift has been confirmed using other analysis approaches by \citet{liang2019} and \citet{sommovigo2022}. 

\begin{table}[!t]
\begin{center}
\caption{Reduced $\chi$2 for different SED templates relative to the available measurements \citep{derossi2018}}\label{tab:chisq}
\begin{tabular}{lll}
Template                                    & z=2-4 &  z=5-7\tablenotemark{a}\\
\tableline\tableline
\citet{rieke2009}, Log(L) = 11.25                    & 1.87  &  6.1  \\
\citet{rieke2009}, log(L) = 12.25 & 2.91 & 4.5  \\
\citet{schreiber2018}  & 3.05  & 3.25 \\
\citet{kirkpatrick2015}  &  4.06  &  -- \\
\citet{magdis2012}  &  2.27  &  -- \\
\citet{chary2001}  &  5.61  &  -- \\
\citet{derossi2018}, Haro 11 & -- &  1.57 \\
\tableline
\end{tabular}
\tablenotetext{a}{Templates that do not anticipate the increase in temperature to this redshift have not been added. We can expect they would be poor fits because their SEDs are too narrow.  This effect can be seen by comparing the two \citet{rieke2009} ones at different luminosity. The higher luminosity template becomes the better fit at high redshift because its peak  is at an appropriate wavelength, but its shape does not fit, as indicated by the high $\chi^2$.}

\end{center}
\end{table}


This change in the SEDs with  redshift can have a significant effect on infrared luminosity estimates using ALMA $\sim$ 1\,mm data, as shown graphically in \citet[][figure~9]{derossi2018}. Another consequence of this difference is that, for a given infrared luminosity, the flux density in the $\sim$ 1 mm  ALMA bands can be substantially lower than predicted by local templates based on observations at optical through mid-infrared wavelengths and used to interpret the ALMA measurements (for examples, see \citealt{williams2024}). 

Resolving this issue is of increasing priority. Dusty galaxies at very high redshift (e.g., z $>$ 5) were  discovered in significant numbers during the previous decade \citep[e.g.,][]{strandet2016, spilker2016, strandet2017, zavala2018}, but only at the highest luminosity ranges,  similar to those studied by \citet{derossi2018}.  The measured properties of these galaxies can be taken as intrinsic {\it only} for the class of very luminous sources observed. ALMA studies are {\bf now} pushing to lower luminosities \citep[e.g.,][]{williams2019, fudamoto2021, zavala2021, xiao2023, hill2024, mitsuhashi2024a, liu2026, faisst2025}; hence, determining the relation between redshift and far infrared SEDs  over a broad range of galaxy 
properties is increasingly important. 
The emerging synergy with JWST \citep[e.g.,][]{derossi2023,ferrara2024,herrera2026} suggests that accurate interpretation of the ALMA results will be of increasing importance, with a particularly puzzling challenge provided by the broad-band SED of the ``little red dot'' (LRD) class of high-$z$ sources \citep[e.g.,][]{chen2025}.  

In this paper, we will probe the galaxy parameters that drive the shape of the far infrared SED. 
We will also make suggestions for interpreting ALMA observations at high-redshift in terms of total infrared luminosity, critical for estimating the rates of obscured star formation.
 In Section~\ref{sec:cause_of_sed_changes}, we discuss the main causes related to changes in SED shapes, highlighting the role of the luminosity density. 
 Section~\ref{sec:modeling_at_very_high_redshift}  expands our analysis to very high redshifts, implementing a model specially designed for the study of early-Universe galaxies. A comparison with previous works is carried out in Section~\ref{sec:comparison_with_other_works}. In Sections~\ref{sec:approach_to_accurate_IR_luminosities} and \ref{sec:estimating_obscured_star_formation}, we analyze approaches to determine IR luminosities and to estimate obscured star formation, respectively.  Finally, our conclusions are summarized in Section~\ref{sec:conclusions}.

\section{Cause of SED Changes}
\label{sec:cause_of_sed_changes}

\begin{figure}[h!]
\begin{center}
\vspace{-0cm}
        \hspace{-0cm} \includegraphics[width=8cm]{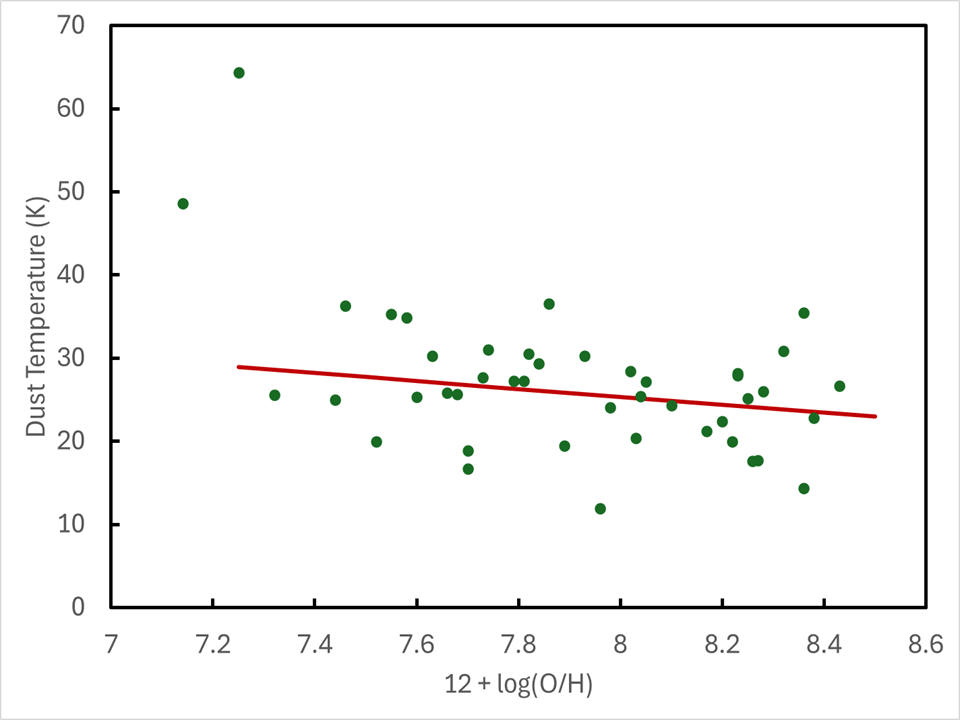}
        \hspace{-0cm} 
\end{center}
\caption{The trend of temperature corresponding to the peak of the far infrared SED vs. galaxy metallicity, from \citet{remy2015}. 
As usual, the oxygen abundance, 12+log(O/H), is used as a proxy for a galaxy's ISM metallicity.
The line is fitted to all points except the two at the lowest metallicity. Further details can be found in \citet{remy2015} and references therein.
}
\label{fig:metals}
\end{figure}

 \citet{derossi2018} concluded that the primary cause of the changes in SED shape in their sample of $z = 5 - 7$ galaxies, i.e., the trend toward higher effective temperature and the broadened SED, is the very high star formation volume density, or equivalently the luminosity volume density in the interstellar medium (ISM) of the galaxies. However, the interplay between metallicity and the properties of the star forming regions has led to  suggestions that low metallicity may also play a role. We have tested this possibility empirically against local galaxies from \citet{remy2015}. As shown in Figure~\ref{fig:metals}, there is little trend in far infrared temperature with metallicity, except for the two lowest metallicity galaxies, I Zw 18 and SBS0335-052.  If these two galaxies are removed, the fit to the points is

\begin{equation}
T/{\rm [K]} = (63.6 \pm 23.8) -(4.8 \pm 3.0) \times (12 + \rm log(O/H))
\end{equation}

\noindent
That is, the relation between metallicity and far infrared SED temperature is marginal, at only 1.6 
standard deviations over the metallicity range relevant to luminous high-redshift galaxies. The apparent dependence on metallicity arises primarily because of its effects on the ionizing output of hot stars. We note that at high-$z$, another reason for an elevated dust temperature may be the rising floor set by the cosmic microwave background (CMB), where radiative processes are not allowed to cool the dust below the CMB temperature \citep[e.g.,][]{schneider2010,safranek2014}, although up to $z \sim 7$, at the luminosities and temperatures we consider, this is a small effect \citep{dacunha2013}.


\begin{figure}[h!]
\begin{center}
\vspace{-0cm}
        \hspace{-0cm} \includegraphics[width=8.5cm]{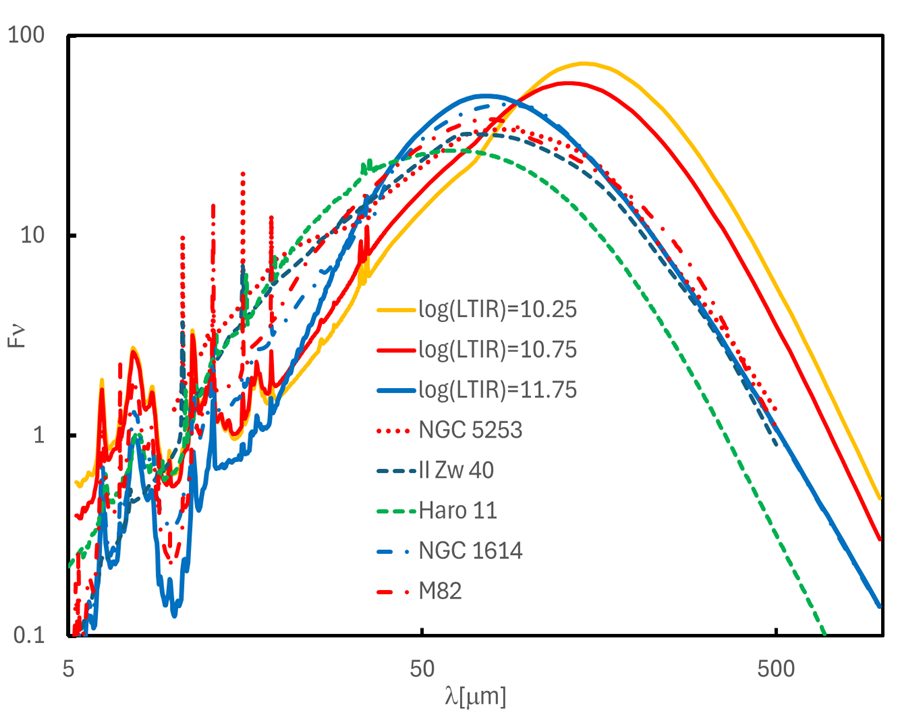}
        \hspace{-0cm} 
\end{center}
\caption{The far infrared SEDs of three galaxies with luminosity densities  $> 10^4~{\rm L}_{\odot}{\rm \,pc}^{-3}$ (NGC 5253, II Zw 40, Haro 11), compared with two with high but somewhat lower luminosity densities (NGC 1614 and M82), and with the average templates for galaxies of similar luminosity. All SEDs are in units of Jy but are normalized to an identical total far infrared luminosity (i.e., from 8 to 1000 $\mu$m).}
\label{fig:density}
\end{figure}\textbf{}

\citet{galliano2011, ciesla2014} and \citet{remy2015}, among others, have pointed out the effects of high starlight intensity on the far infrared SEDs of galaxies, broadening and shifting them to warmer temperatures. In general, this intensity is difficult to estimate directly from observations and it has been deduced from the characteristics of the infrared SED \citep[e.g.,][]{galliano2011}. As a result, using such estimates of starlight intensity to deduce characteristics of the infrared SED of an individual galaxy is likely to be circular in reasoning.

  Direct estimates of the luminosity density are feasible for only a few individual very luminous galaxies, due to the need to understand the structure of the star forming region in three dimensions. We focus on fiducial cases where high luminosity densities can be determined with some confidence.  In these cases, where the star-forming regions are heavily dust-embedded, we use the infrared luminosity as a proxy for the luminosity in UV photons.  This is consistent with the UV photons being almost entirely absorbed by the interstellar dust in heavily obscured star forming regions  \citep[e.g.,][]{alonso2006}. We show some examples in Figure~\ref{fig:density}:

\begin{itemize}

\item{Haro 11 has a complex irregular structure believed to result from an ongoing merger, with three prominent knots in the visible. From infrared imaging,  the star formation is largely concentrated in Knot B \citep{lyu2016}. \citet{adamo2010} present HST optical-to-red images that indicate a diameter of no more than 200 pc for this feature, although the result might be affected by extinction. \citet{lyu2016} show a Spitzer/MIPS 24 $\mu$m image where Knot B is completely unresolved, placing an upper limit of $\sim$ 2$''$ on its diameter, whereas \citet{ostlin2015} present a NICMOS 1.6 $\mu$m image indicating Knot B is $\lesssim$ 1$''$, 
$\lesssim$ 400 pc, in diameter. Following \citet{derossi2018}, we assume a radius of 140 pc and a luminosity of $2 \times 10^{11}$ L$_\odot$ \citep{adamo2010}, finding that the average luminosity density is $\sim 1.5 \times 10^4$ L$_\odot$ pc$^{-3}$, presumably with even higher values in many sub-regions.
}

\item{NGC 5253 is dominated by small, $<$ 1\,pc, star clusters within a diameter of 0.5$''$ \citep{smith2020}. The luminosity is $1.1 \times 10^9$ L$_\odot$ \citep{sanders2003}. Taking the entire 0.5$''$ region to generate the infrared output, the average luminosity density is $\sim 3 \times 10^5$ L$_\odot$ pc$^{-3}$
.} 

\item{ In the case of II Zw 40, the size is about 20 $\times$ 30 pc \citep{kepley2014}, while the luminosity is $1.9 \times 10^9$ L$_\odot$ \citep{beck2002}, yielding an average luminosity density of $\sim 3 \times 10^4$ L$_\odot$ pc$^{-3}$.}

\item{The star forming region in NGC 1614 is a donut with a volume of about $6 \times 10^7$ pc$^{3}$ \citep{pereira2015} and a luminosity of 
$4 \times 10^{11}$ L$_\odot$ \citep{sanders2003}, for an average luminosity density of $\sim 6 \times 10^3$ L$_\odot$ pc$^{-3}$.}

\item{The starburst region in M82 can be taken to be a disk, which is close to edge-on since the starburst-powered wind is in the plane of the sky. We estimate it to be  210 pc in radius and 70 pc in height \citep{kronberg1981}. With a luminosity of $6 \times 10^{10}$ L$_\odot$  \citep{sanders2003}, this yields an average luminosity density of about $6 \times 10^3$ L$_\odot$ pc$^{-3}$. }

\end{itemize}

We will compare these galaxies with templates of ``normal'' local infrared galaxies from \citet{rieke2009}. These templates are  also applicable at higher redshift. The Log($L_{\rm TIR}$$) = 11.25$ template is a good fit to high redshift galaxies of LIRG or ULIRG luminosity up to $z \sim 4$ \citep{derossi2018}, as well as (obviously) being representative of local LIRGs.

In Figure~\ref{fig:density}, for the same total infrared energy output, the three  galaxies with the highest luminosity densities (including Haro 11) have SEDs  above the average templates from \citet{rieke2009} by factors of 
$\sim 5$ at mid-infrared wavelengths. The two galaxies plotted with high but not maximal luminosity densities, NGC 1614 and M82, also fall above the appropriate standard templates in the mid infrared, log(LTIR) = 11.75 and log(LTIR)=10.75. This demonstrates that the extreme behavior of Haro 11 is not unique, but is likely just a result of the extremely high luminosity density in its dominant star-forming region.

This behavior is frequently seen in the most luminous galaxies at high redshift. For example, figure~7 of \citet{spilker2016} shows luminosity densities of $10^3 - 10^4$ L$_\odot$\,pc$^{-3}$ for high redshift luminous galaxies, even when assuming that the luminosity is uniformly distributed in a sphere. These values are presumably higher with a more realistic spatial structure. 
This is likely to be the basic cause for the far infrared SEDs of these galaxies more closely resembling the spectrum of Haro 11 rather than local templates of similar luminosity \citep{derossi2018}.

\section{Modeling at very high redshift}
\label{sec:modeling_at_very_high_redshift}

The preceding sections have focused on nearby galaxies and the effect of very high luminosity densities on their mid-infrared SEDs. An extensive set of models illustrating this behavior is presented in \citet{galliano2011}, aimed toward a spatial analysis of the emission of the Large Magellanic Cloud.  We now expand these results to very high redshift, using a detailed model for early-Universe galaxies. Our framework is based on that described in 
\citet{derossi2017}, and further developed in \citet{derossi2019,derossi2023}. Here, we employ an improved version of those models adapted to the specific problem of individual observed galaxies \citep{derossi2018} to probe the behavior of the far infrared SEDs. 
 In particular, for this work, we assume a default radius $R=1.5$~kpc for the ISM component (Section~\ref{sec:default}).   Unless otherwise specified, $R$ will be considered a fixed parameter in the model; hence, differing SFRs correspond to different SFR (luminosity) densities.  Our adopted galaxy radius is representative of luminous and massive sources at $z \sim 4 - 8$. For example, \citet{ormerod2024} find $\left< r_e \right>$  $\sim$ 1.2 kpc for galaxies of $\log(M) >$ 9.5 at z $\sim$ 6, and \citet{allen2025} find $\left< r_e \right> >$ 1 kpc for $\log(M) > 9.2$. The delensed models of luminous submm galaxies  of $4 < z < 6$ by \citet{spilker2016} have  $\left< r_e \right> \sim $ 0.9 kpc (omitting the anomalously small galaxy SPT2319-55).

The  galaxies typically observed at cosmic dawn with JWST  tend to be smaller \citep[e.g.,][]{morishita2024,ono2025} and of lower luminosity and mostly beyond the reach of ALMA continuum detections \citep[e.g.,][]{hill2024,liu2026}. The effects of varying sizes, among other key model parameters, will be discussed in Section~\ref{sec:variations}.   However, galaxies as small as the typical JWST-discovered ones are not included in our analysis.

\subsection{Underlying assumptions to models}
\label{sec:model}

A model galaxy consists of a central stellar cluster, surrounded by a mixed phase of dust and gas. For simplicity, we adopt a homogeneous and spherically symmetric ISM component.

\subsubsection{Stellar emission}

We assume a constant SFR lasting for a period of time $T$, which ends at a given reference redshift $z=z_{\rm ref}$. 
Thus, the stellar mass $M_\star$ of a system, at $z=z_{\rm ref}$, is $M_* = {\rm SFR} \times T$. 
Our fiducial model adopts $z_{\rm ref}= 5$, but  the general trends explored in this work do not depend on this choice.
Assuming ${\rm SFR =} 30 ~{\rm M}_\odot ~ {\rm yr}^{-1}$ and $M_* = 2 \times 10^9 ~{\rm M}_\odot$ for a Haro 11 model, we set $T=67$ Myr.
In addition to ${\rm SFR} = 30~{\rm M}_\odot ~ {\rm yr}^{-1}$, for  comparison we considered the cases SFR: 5, 10, and 20 ~${\rm M}_\odot ~ {\rm yr}^{-1}$ (preserving the parameter $T=67$ Myr).  

The location of our model sources ($5~{\rm M}_\odot~ {\rm yr}^{-1} \le {\rm SFR} \le  30~{\rm M}_\odot~ {\rm yr}^{-1}$)  in the ${\rm SFR}-{M_*}$ plane is consistent with observations at $z~\sim5$,   with model galaxies lying close to the main sequence (MS) reported in literature \citep[e.g.][]{Santini2017, Rinaldi2022, williams2024, DiCesare2026}.

To generate stellar SEDs, we use Yggdrasil model grids \citep{zackrisson2011}. Yggdrasil is a population synthesis code specifically designed for describing the behavior of high-$z$ galaxies, though it can also be used for modelling galaxy SEDs at lower $z$.  Yggdrasil provides the SEDs for simple stellar populations with different ages and star-formation histories, considering stellar metallicities ranging from zero to supersolar.
We construct the total stellar SED of a model galaxy by adding the SEDs of different single-age stellar populations. Following \citet{derossi2017}, we adopted grids corresponding to the lowest stellar metallicity available for Population II stars ($Z_* = 0.0004$) and a Kroupa initial mass funtion in the interval $0.1 - 100~{\rm M}_\sun $.
The integrated specific luminosity associated with the stellar component of a model galaxy is calculated as:

\begin{equation}
	L_{\nu, *, {\rm tot}} = \sum_i L_{\nu, *}({\tau}_i), 
\end{equation}

\noindent where $L_{\nu, *}({\tau}_i)$ is the specific luminosity of a stellar population with age ${\tau}_i$.

\subsubsection{Gas-phase metallicity}

In the Local Universe, there is a well-known correlation between the stellar mass and gas-phase metallicity ($Z_{\rm gas}$) of galaxies
(MZR) \citep{tremonti2004}. Different studies suggest that the MZR extends towards higher $z$, although with a lower normalization \citep[e.g.,][]{maiolino2019}.  

\citet{sarkar2025} 
estimated the MZR at $z=4-10$ with JWST/NIRSpec measurements.  They obtained the MZR slope ($\gamma$) and normalization ($Z_{10}$) in different $z$ intervals by fitting observational data with the expression:

\begin{equation}
	\label{eq:MZR}
	12+ \log ({\rm O/H})= \gamma \log\Big( \frac{M_*}{10^{10}{\rm M}_\odot}\Big) + Z_{10}.
\end{equation}

\noindent 
In their Table~1, $\gamma$ and $Z_{10}$ are reported for different $z$ bins. At $z=4-6$, for instance, $\gamma \approx 0.28$ and $Z_{10} \approx 8.37$.

Following the usual custom,  Equ.~\ref{eq:MZR} uses the oxygen abundance as a proxy for metallicity.
We converted
absolute oxygen abundances to the metallicities relative to solar 
assuming the solar abundances adopted in \citet{sarkar2025}, which correspond to the values reported by \citet{anders1989}.  Then, an offset was applied to the obtained metallicities so that they are consistent with our adopted solar value (i.e., $Z_{\odot} = 0.0142$).
Implementing these corrections, the relation between 
$\log(Z_{\rm gas}/Z_\odot)$ and $M_*$ can be approximated as:

\begin{equation}
        \label{eq:MZR2}
        \log(Z_{\rm gas}/Z_\odot)= \gamma \log\Big( \frac{M_*}{10^{10}{\rm M}_\odot}\Big) + Z_{10} -8.80583.
\end{equation}
\noindent
We use Equ.~\ref{eq:MZR2} to estimate the gas-phase metallicity of a model galaxy with a stellar mass $M_*$.

\subsubsection{Dust-to-metal ratio}

Scaling relations associated with the dust-to-metal ratio ($D/M$) of high-$z$ galaxies are still not robustly constrained.
In this work, we implement the linear regression fit, reported by \citet{popping2022}, for the observed relation between $D/M$ and 12+log(O/H):

\begin{equation}
        \label{eq:DM_Mstar_relation}
\log(D/M) = 0.32 \times [{\rm 12+\log(O/H)}] - 3.28.
\end{equation}
\noindent
\citet{popping2022} found that such a relation seems to not evolve significantly, at least in the range $0<z<5$.  
By applying Equ.~\ref{eq:DM_Mstar_relation}, we derive $D/M$
for a model galaxy with a given ${\rm 12+\log(O/H)}$. Oxygen abundances are estimated from  Equ.~\ref{eq:MZR}, using $M_*$.

\subsubsection{Gas mass}

Gas masses ($M_{\rm gas}$) of high-$z$ sources are difficult to determine  accurately. 
To roughly estimate the  dependence of $M_{\rm gas}$ on $M_*$, we used the fitting relations given in \citet{scoville2016}, which describe the relation between molecular gas fraction and $M_*$ for different $z$. 
We consider the special case of galaxies on the MS.  
In this case, the molecular gas mass fraction
can be estimated from the following expression:

\begin{equation}
        \label{eq:fgas_Mstar_relation}
	f_{\rm gas} = \frac{M_{\rm mol}}{M_{\rm mol}+M_*} = 0.30 \Big( \frac{M_*}{10^{11}{\rm M}_\odot}\Big)^{-0.02} \Big( \frac{1+z}{3}\Big)^{0.44}.
\end{equation}
\noindent
We estimated $M_{\rm gas}$ using
Equ.~\ref{eq:fgas_Mstar_relation}, assuming that $M_{\rm gas} \sim M_{\rm mol}$ at high $z$.  For a MS galaxy, a gas fraction of $\sim 0.4$ is obtained at $z\sim 5$, with no significant variations with $M_*$ and $z$.  To consider galaxies off the main sequence, a factor $({\rm sSFR} / {\rm sSFR}_{\rm MS})^{0.32}$ should be added to  Equ.~\ref{eq:fgas_Mstar_relation}, where sSFR is the specific SFR of the galaxy and ${\rm sSFR}_{\rm MS}$ that of a MS galaxy of similar $M_*$.  Therefore, galaxies above (below) the MS by one order of magnitude
would have higher (lower) gas fractions by a factor of $\sim 2$ ($\sim 0.5$). 
In Section~\ref{sec:variations}, we will evaluate the impact on the SEDs of assuming different gas fractions.



\subsubsection{Dust emission}
\label{sec:dust_emission}

We estimate dust emission from a dominant silicate-rich dust component (80\%), combined with a low level of emission from carbon-based dust (20\%). Earlier on, it was thought that dust in the very early Universe would be silicate rich with very little carbon \citep[e.g.,][]{cherchneff2010}. However, carbon emission lines have been detected even in galaxies at $z > 12$ \citep{deugenio2024, carniani2025, naidu2025}, and carbonaceous dust grains have also been found out to $z \sim 8$ \citep{witstok2023}. It appears that the early production of carbonaceous dust is multi-faceted, combining the output of AGB stars and supernovae \citep[e.g.,][]{nanni2025,chiaki2025}.  In any case, a realistic model of high-redshift interstellar dust must include a significant level of carbon.

 For the silicon-based dust, our fiducial model adopts the UM-D-20 chemistry described in \citet[][]{cherchneff2010} and the so-called standard size distribution of dust grains implemented in \citet{ji2014}.\footnote{ Other silicon-based dust chemical compositions are presented in \citet{cherchneff2010}. Using a different silicon-based chemistry does not result in  significant changes in the general shape and normalization of the model SEDs; only the specific spectral features are affected \citep[e.g.,][]{derossi2018}.  This is consistent with some heuristic experiments that we performed, which additionally show that varying the lower grain size limit has
very little effect, and varying the upper limit has only a very modest effect on the resulting SED. Changing the slope of grain sizes within the distribution also has
virtually no effect, if it is done within the usual bounds for the distribution of interstellar dust.
}
Following \citet{derossi2018}, the infrared SED expected for amorphous carbon has been derived using optical constants
from K. Misselt (2018, private communication), based on those
of \citet{zubko1996}. The model does not include stochastically heated very small grains and hence does not include PAH emission. The aforementioned assumptions have been shown to be adequate for matching the  far-infrared SED of Haro 11, a useful analog to luminous, high-redshift galaxies (see the discussion in \citealt{derossi2018}).

Dust temperature ($T_{\rm d}$) is obtained by assuming thermal equilibrium, considering the heating of dust grains by the stellar source and the CMB at the corresponding redshift. A dust sublimation temperature of 1500 K and 2000 K is assumed for Si- and C-based dust grains, respectively.  The dust emissivity is estimated by applying Kirchhoff’s law to the resulting $T_{\rm d}$-profile.  A detailed description of the methodology used for these calculations can be found in \citet{derossi2017}.

 By implementing the prescriptions summarized above, our model can predict the SEDs corresponding to dust emission for sources with different SFRs at high $z$. To improve the model fit to long-wavelength observations, we apply a $\beta$ term analogous to that used in modified blackbody fits. In our case, we multiply the model predictions for $\lambda > 150\, \mu$m by 

\begin{equation}
   \left( \frac{150~ \mu {\rm m}}{\lambda}\right)^\beta \rm{\ ,}
\end{equation}
where $\beta = 0.57$.
\noindent
Because the model SED already falls faster than a blackbody toward long wavelengths, the adopted value of $\beta$ is smaller than those used with modified blackbodies. We next summarize the main model parameters adopted in our fiducial model.

\subsection{Fiducial Model}
\label{sec:default}

As a foundation, we have calculated a set of default models based on the properties of Haro 11, which is a good reference local analog of luminous, high-$z$ galaxies.  The input parameters are discussed above; we summarize them here:

\begin{itemize}
        \item  Reference redshift: $z_{\rm ref}=5$.
        \item  Radius of the ISM (gas and dust) component: $R=$ 1.5 kpc (consistent with an Haro 11 model; see \citealt{derossi2018}). Assuming a fixed galaxy size maps the SFR directly to the luminosity density to allow a simple illustration of the governing physics underlying the far infrared SED. 
	    \item  Star formation rate: SFR/[${\rm M}_\odot ~ {\rm yr}^{-1}$] $=$ 5.0, 10.0, 20.0 and 30.0 (similiar to Haro 11).   
         \item  Period for star formation: $T=$ 67 Myr (considering an SFR= 30.0 ${\rm M}_\odot ~ {\rm yr}^{-1}$, this value of $T$ correponds to a model for Haro~11, as discussed previously).
        \item  Gas-phase metallicity ($Z_{\rm gas}$): given by Equ.~\ref{eq:MZR2}.
        \item  Dust-to-metal ratio ($D/M$): given by Equ.~\ref{eq:DM_Mstar_relation}.
        \item  Gas mass ($M_{\rm gas}$): given by Equ.~\ref{eq:fgas_Mstar_relation}.
	   \item  Dust composition: 20\% and 80\% of the total emissivity contributed by carbon and silicate, respectively.
\end{itemize}
\noindent

 In the case of SFR$=20-30~{\rm M}_\odot ~ {\rm yr}^{-1}$, Equ.~\ref{eq:MZR2}-\ref{eq:fgas_Mstar_relation} predict $Z_{\rm gas}$, $D/M$ and $M_{\rm gas}$ comparable with estimates for Knot B in Haro 11 \citep[e.g.][]{Guseva2012, James2013, ostlin2015}. We checked that variations in such parameters (preserving the SFR value) only lead to changes in the SED normalization, bearing no implications to our analysis regarding the SED shape (see, also, \citealt{derossi2018b}; Fig.~\ref{fig:SEDs_3}, bottom left panel).

\begin{figure}[]
\begin{center}
\vspace{-0.0cm}
        \hspace{-0cm} \includegraphics[width=8.5cm]{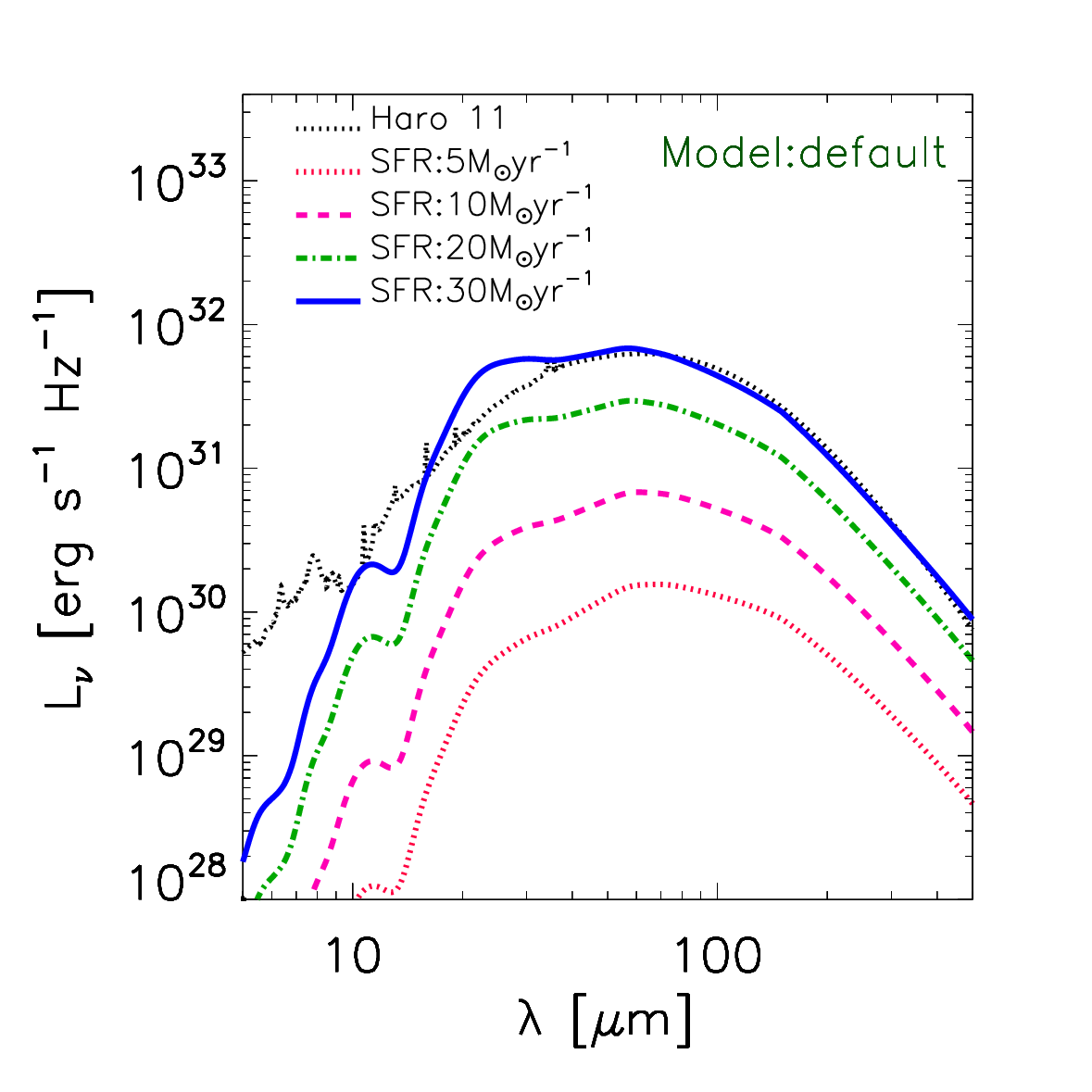}
        \hspace{-0cm} 
\end{center}
\caption{
        Dust specific luminosity obtained from our default  model, assuming different SFRs. For comparison, the SED of Haro 11 is shown with a dotted black line.  In the default model, $Z_{\rm gas}$, $D/M$ and $M_{\rm gas}$ are derived from empirical scaling relations (see Equ.~\ref{eq:MZR2}, Equ.~\ref{eq:DM_Mstar_relation}, and Equ.~\ref{eq:fgas_Mstar_relation}, respectively). }
\label{fig:SEDs_1}
\end{figure}

In Fig.~\ref{fig:SEDs_1}, we show the predictions derived from our default model. We explored different SFRs, corresponding to different luminosity densities in the ISM of the model galaxy. Encouragingly,  at ${\rm SFR} = 20 - 30~{\rm M}_\odot {\rm \,yr}^{-1}$, the model describes the general behavior of Haro 11, at least at $\lambda \gtrsim 15~{\mu}{\rm m}$. As a consequence of our adopted scaling relations (Section~\ref{sec:model}),
within our model a decrease in the SFR implies a decrease in $M_*$, $M_{\rm gas}$, $Z_{\rm gas}$ and $D/M$.  As the SFR and luminosity decrease, the SEDs move to longer $\lambda$. 

In Fig.~\ref{fig:SEDs_2},  $M_{\rm gas}$, $Z_{\rm gas}$ and $D/M$ are fixed at their values at ${\rm SFR} = 30~{\rm M}_\odot {\rm \,yr}^{-1}$. 
This figure illustrates more clearly how   changes in the SFR are responsible for the aforementioned SED trends.  As expected, when only the SFR (or equivalently $M_*$) is varied (preserving the values of all other parameters), the changes in the total luminosity are weaker. However, the figure makes it clear that the SEDs corresponding to higher SFR (higher luminosity density) are both shifted to shorter wavelengths and are also broader than those for lower SFRs.

\subsection{Discussion of the default model}

 Figure~\ref{fig:summary} compares our default model with various templates and with the relevant measurements at high redshift. The figure shows that our model is a good fit to the expected far infrared SED, but falls far below the measurements at $\lambda \le 15$
$\mu$m, as a consequence of our omission of the sources of emission there, e.g. tiny grains and PAH molecules.  For wavelengths $\ge$ 15 $\mu$m, both our model and the SED of Haro 11 fit the data acceptably well (the reduced $chi^2$ for Haro 11 is 1.57 \citep{derossi2018} while for our model it is 2.3). In both cases, the quoted errors on the individual measurements may be a bit optimistic  due to intrinsic variations from galaxy to galaxy; the region between 15 and 100 $\mu$m has a number of points that appear to be too high or too low by more than indicated by the nominal errors. No single SED can simultaneously fit all these points.  

The emission short of 50 $\mu$m is complex to model: (1) it begins to be dominated by single-photon heating of dust grains; (2) at $\lambda < $ 25 $\mu$m, additional complexity arises from silicate spectral features that can be in emission or absorption; and (3) at $\lambda \le$ 15 $\mu$m, an additional component due to PAH molecules comes into play. Because we are focusing on the far infrared, our model need  not include this region with its complexities. 

Templates based on local galaxies with log(L) = 11.25 and 12.25  are too narrow in the far infrared and thus cannot provide good fits, with $\chi^2$ values of 6.1 and 4.5 respectively \citep{derossi2018}. The new data from ALMA agree with the trends in the data summarized in \citet{derossi2018} but do not extend to short enough rest wavelengths for a demanding test (and the cases plotted are among the best available in this regard). The single known case at $z > 8$ shows a different behavior, and dusty SEDs at this and higher redshifts appear to be very rare \citep{laporte2017,bakx2025}.

\begin{figure}[]
\begin{center}
\vspace{-0.0cm}
        \hspace{-0cm} \includegraphics[width=8.5cm]{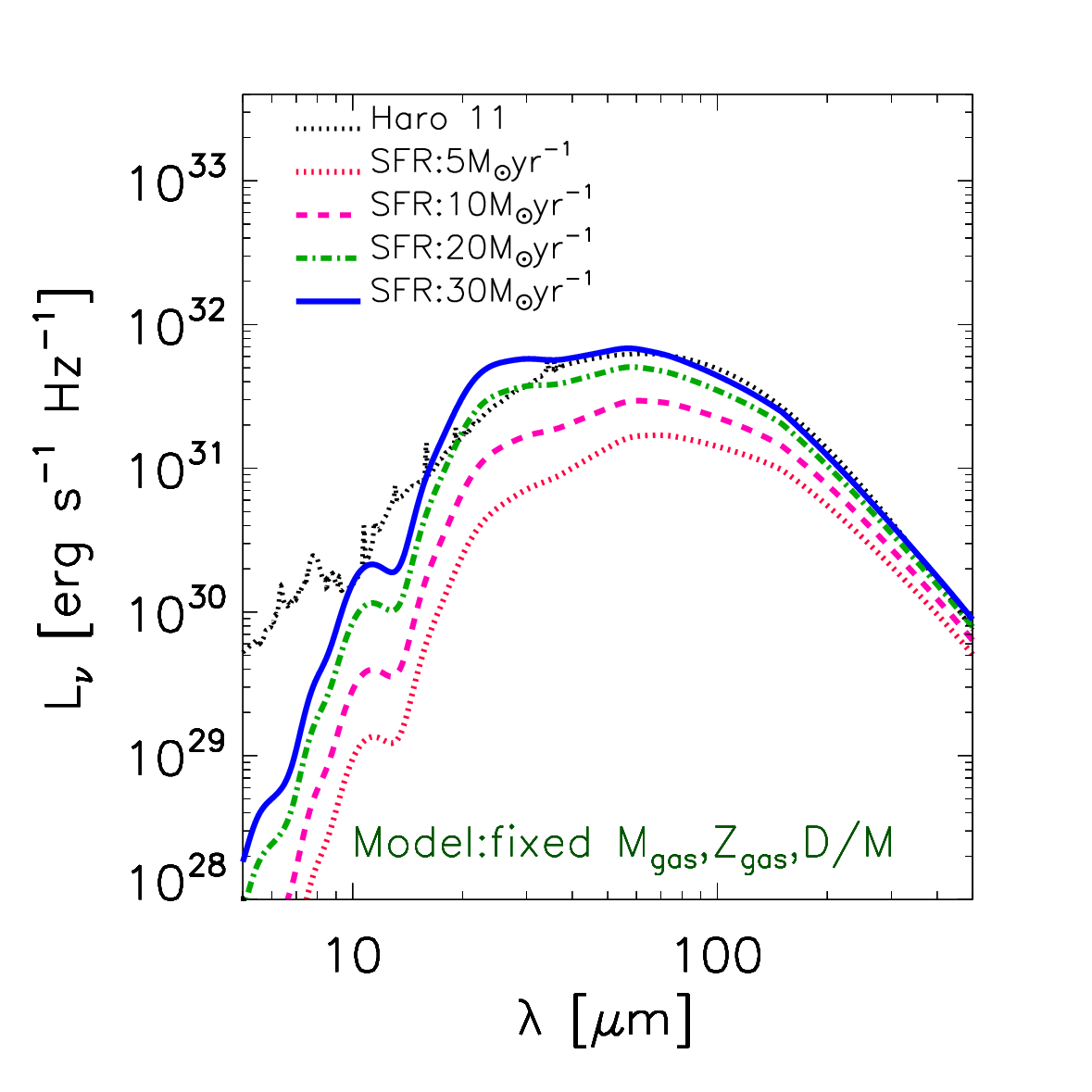}
\end{center}
\caption{
        Similar to Figure~\ref{fig:SEDs_1}, except we have  fixed the parameters $Z_{\rm gas}$, $D/M$ and $M_{\rm gas}$, at their default values at ${\rm SFR} = 30~{\rm M}_\odot {\rm \,yr}^{-1}$.  Hence, the different SEDs correspond to variations in SFR only, i.e., solely to variations in the luminosity density, preserving the values of all the remaining model parameters.}
\label{fig:SEDs_2}
\end{figure}

\begin{figure}[h!]
\begin{center}
\vspace{-0cm}
        \hspace{-0cm} \includegraphics[width=9cm]{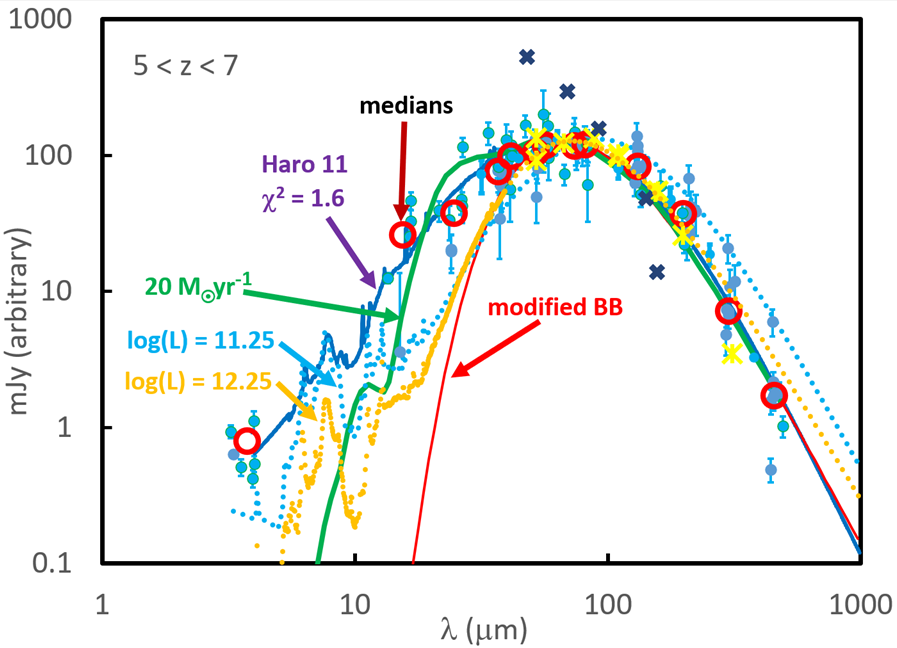}
        \hspace{-0cm} 
\end{center}
\caption{Far infrared SED summary. The figure is based on Figure 7 of \citet{derossi2018}. The blue points with error bars are measurements primarily from Herschel and the South Pole Telescope; the large open circles are medians of those points. The  log(L)=11.25 and 12.25 templates \citep{rieke2009} are in light blue and orange, the Haro 11 SED is in blue and the result of our model for SFR = 20 M$_\odot$ yr$^{-1}$ in green. The modified blackbody has T = 47K and $\beta$ = 1.6. We have also added ALMA measurements that extend to $\sim$ 50 $\mu$m, i.e. for A1689-zD1 from \citet{mitsuhashi2024b} and HZ 10 from \citet{algera2026}, both in yellow X's. The dark blue X's are for the warm far infrared spectrum of the galaxy at z = 8.3 from \citet{bakx2025}. }
\label{fig:summary}
\end{figure}

\begin{figure*}[h!]
\begin{center}
\vspace{-0cm}
        \hspace{-0cm} \includegraphics[width=8cm]{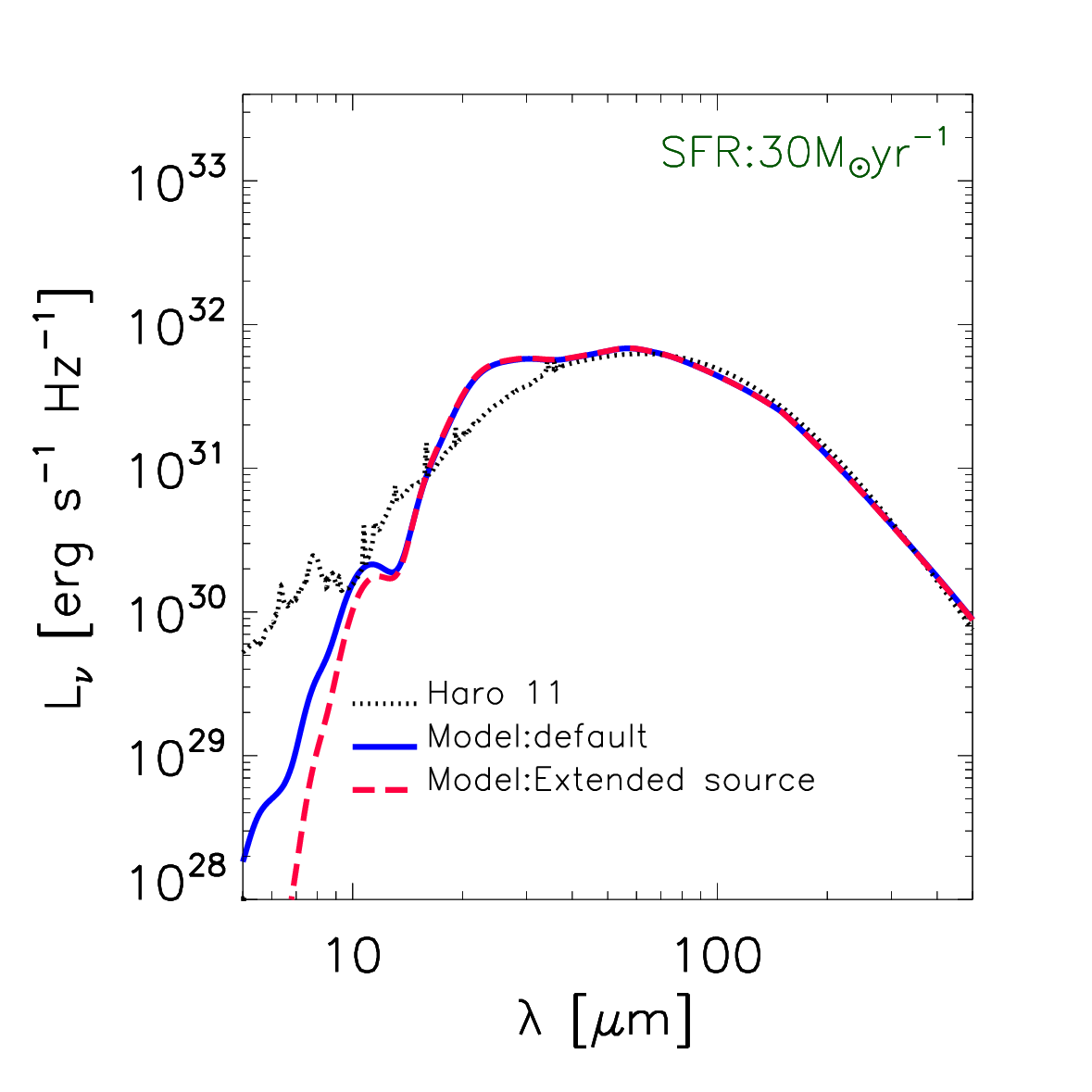}
        \hspace{-0cm} \includegraphics[width=8cm]{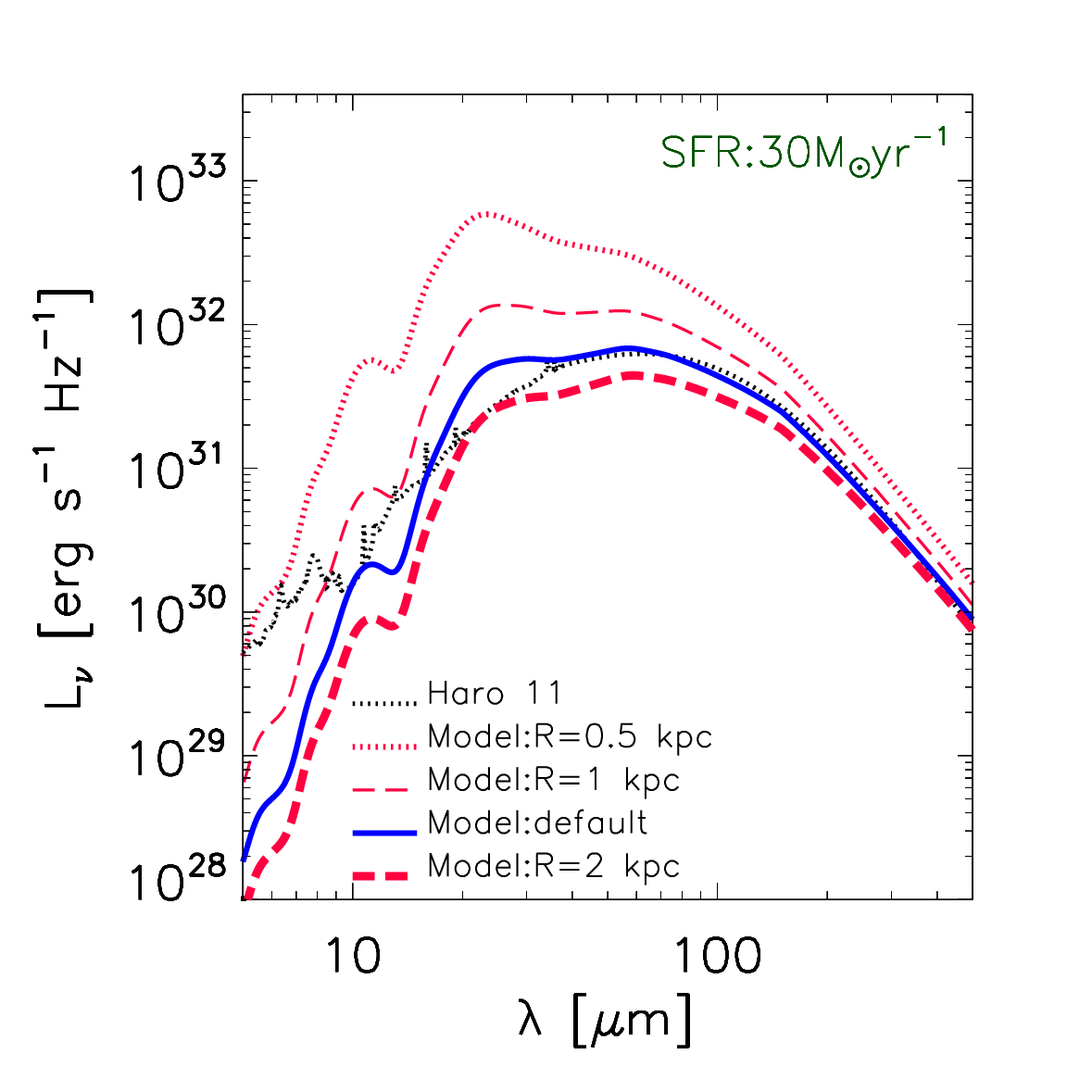}\\\vspace{-0.0cm}
        \hspace{-0cm} \includegraphics[width=8cm]{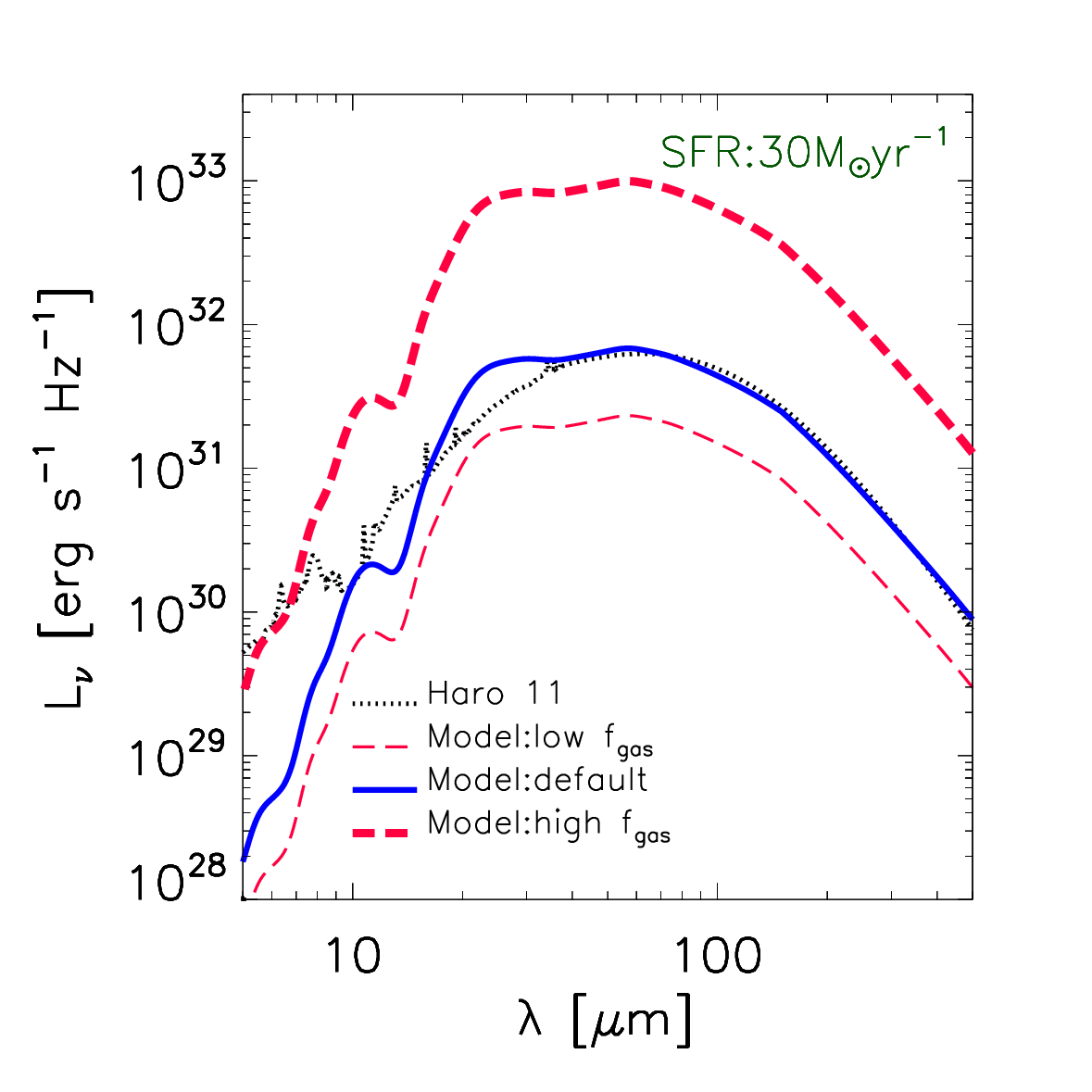}
        \hspace{-0cm} \includegraphics[width=8cm]{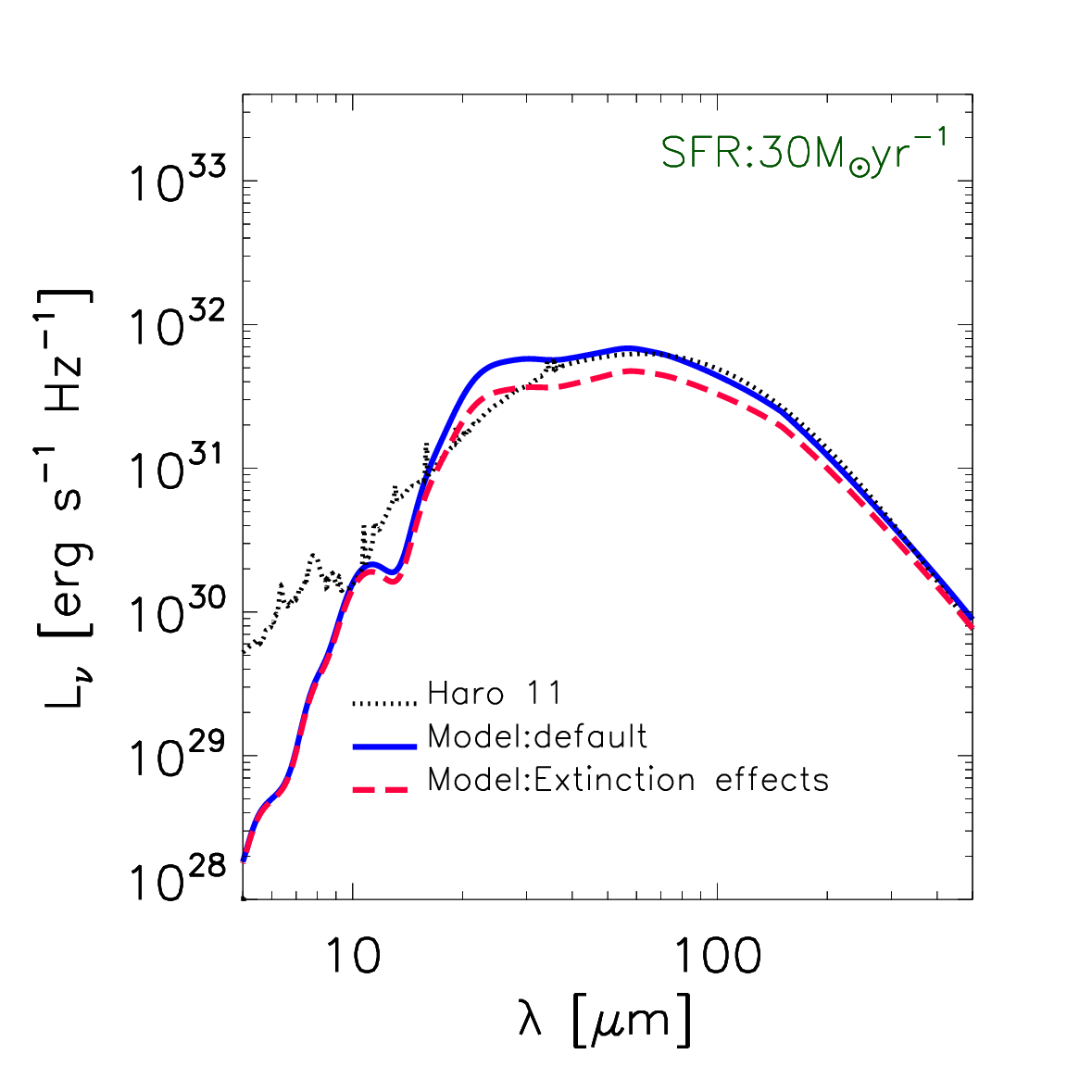}
        \vspace{-0cm}
\end{center}
\caption{
Comparison between the default model SED (blue solid line), corresponding to a compact stellar source, a radius $R=1.5$ kpc for the ISM component, a gas fraction based on main-sequence galaxies, and no extinction effects (Section~\ref{sec:default}), and model SEDs derived when assuming an extended stellar source (top left panel, red dashed line), different radii ($R$) for the ISM (top right panel, red dashed and dotted lines, as shown in the legend), different gas fractions (bottom left panel, red dashed lines) and extinction effects (bottom right panel, red dashed line). For a more detailed description, see the text.
	For comparison, the SED of Haro 11 is shown with a dotted black line.}
\label{fig:SEDs_3}
\end{figure*}

We have carried out side calculations to test some aspects of our model. Pure silicate dust without changing the grain size distribution generally yields much stronger silicate emission features than seen in Haro 11. As described in \citet{derossi2018}, adding a modest amount of carbon reduces these features for some of the compositions, leading to a selection of UM-D-20 silicate chemistry with 20\% carbon there, which we also use in this paper.  
The models are degenerate and modest variations in properties such as the slope of the grain size distribution can be accommodated by varying other model parameters. Fortunately, our study depends on the {\it relative} behavior when holding the  parameters constant, except only  for the one of interest, (luminosity density),  as illustrated in Fig.~\ref{fig:SEDs_2}. The effect on the SED of luminosity density is the main result of our modeling.

\subsection{Variations in the model prescriptions}
\label{sec:variations}

Fig.~\ref{fig:SEDs_3} evaluates the effects of changing key model prescriptions.
In the top left panel, we compare the predictions of the default model (blue solid line) with the SED corresponding to an  extended stellar source (red dashed line). The default model considers a compact (point-like) stellar source, assuming a
single, central luminosity source, while the extended configuration arises from
sources randomly distributed within a 100~pc radius.  The cases for an extended star cluster versus a point-like central source converge for wavelengths
$> 10 \mu {\rm m}$, such that the distribution of heating sources to first order has relatively little effect on the predicted FIR SED.  Similar results can be found in \citet[][see their fig.~2]{derossi2018}.

In the top right panel, we consider again a single, central stellar source but, in this case, we vary the radius ($R$) of the ISM (gas and dust) component.  Note that the masses of the gas and dust components are preserved; only their sizes (and, hence, their mass densities) are changed. We see that a smaller $R$ leads to a higher total infrared luminosity because of the increase in dust temperatures, which in turn is a consequence of the higher mass densities reached by the ISM in the case of smaller galaxies. Dust grains are more concentrated around the central stellar source, being more efficiently heated.  Consequently, the dust emissivity increases, specially at shorter wavelengths.

As previously mentioned,  gas masses at high $z$ are very difficult to estimate accurately, rendering $M_{\rm gas}$ a highly uncertain parameter in our modeling.  In particular, Equ.~\ref{eq:fgas_Mstar_relation} was obtained from the analysis of a sample of massive galaxies ($M_* \gtrsim 10^{10}~{\rm M}_{\odot}$) on the main sequence.
At $z\approx 5$, it predicts a gas fraction of $\sim 0.4$, with very modest variations with mass.
As noted above, for galaxies off the main sequence, a factor $({\rm sSFR} / {\rm sSFR}_{\rm MS})^{0.32}$ should be added to  Equ.~\ref{eq:fgas_Mstar_relation}.  Given the uncertainties involved in the estimates of $M_{\rm gas}$, we compare the predictions of our default model with two extreme models for galaxies above (${\rm sSFR} / {\rm sSFR}_{\rm MS} = 10$) and below (${\rm sSFR} / {\rm sSFR}_{\rm MS} = 0.1$) the main sequence. The gas fractions for the latter models are about 0.9 and 0.2, respectively. In the bottom left panel of Fig.~\ref{fig:SEDs_3}, we see that an increase of the gas fraction generates an increase of the dust luminosity. This is expected because, in our dust model, higher gas masses imply higher dust masses, too. In spite of the increase of the SED normalization with the increase of the gas fraction, its shape does not seem to depend significantly on it.


Finally, in the bottom right panel of Fig.~\ref{fig:SEDs_3}, we compare the results of our default model, which does not include extinction effects (thus assuming an optically thin medium), with a model that implements them.  We model dust extinction assuming that stellar radiation is attenuated by a factor $e^{- {\tau}_{\nu}}$.  Taking into account that our model adopts a constant dust mass density ${\rho}_{\rm d}$ and a minimum radius $r_{\rm d} = 5~{\rm pc}$ for the dust component \citep[][]{derossi2017}, the frequency-dependent optical depth is given by
${\tau}_{\nu} = {\kappa}_{\nu} (r - r_{\rm d}) {\rho}_{\rm d}$, where ${\kappa}_{\nu}$ is the frequency-dependent dust opacity, and $r$ denotes the distance to the stellar source.
The contribution of both C- and Si-based dust was considered for the computation of the optical depth.
According to Fig.~\ref{fig:SEDs_3}, extinction effects are very insignificant for our analyzed sources.  Furthermore, we verified that extinction effects are reduced for galaxies with lower SFRs.

These results imply that the prediction of relatively  high temperatures for galaxies with high luminosity densities is reasonably robust.

\subsection{The Eddington Limit}

As discussed previously, our default model assumes $R=1.5$ kpc, consistent with new findings regarding the size\--mass relation at $z \sim 3-9$ \citep[e.g.,][]{allen2025}. Recent results suggest smaller sizes (effective radius $\lesssim 0.5$ kpc) for lower mass galaxies at cosmic dawn \citep[e.g.,][]{finkelstein2023}.  If we compare our default SED with one obtained adopting a conservative $R=0.5$ kpc, large discrepancies can be seen at $\lambda \lesssim 50~{\mu {\rm m}}$, where the luminosities associated with $R=0.5$ kpc can be more than one order of magnitude higher than the default case.\footnote{ It is worth highlighting that only the parameter $R$ is varied in the top right panel of Fig.~\ref{fig:SEDs_3}. The redshift is fixed at $z=5$ (default). }
An additional increase in luminosity is expected at higher $z$ due to the increase of the temperature floor imposed by the CMB background.

However, the most luminous and compact star forming galaxies fall close to the Eddington star formation limit \citep{thompson2005,murray2005}, where the pressure due to escaping radiation ejects the dust and hence the gas and limits the fuel for continued star formation. As an example, for component B of Haro 11 we take a gas fraction $f_g = 0.4$ and the velocity dispersion to be 24 km s$^{-1}$  \citep{gao2022}; using the formulation in \citet{murray2005} the limiting Eddington luminosity is $\sim 6 \times 10^{11}$ L$_\odot$. The measured luminosity is nearly as large, $\sim 2 \times 10^{11}$ L$_\odot$. 
The measurement is an average luminosity, whereas the star formation will operate on the individual H II regions and molecular cloud ``hot spots,'' each of which is likely to be Eddington-limited.  
This provides an explanation for why even broader SEDs  (compared with Haro 11) in Figure~\ref{fig:SEDs_3} (upper right) 
are seldom if ever seen.

\section{Comparison with Other Works}
\label{sec:comparison_with_other_works}


A number of other studies have focused on the cause of the temperature increase of the far infrared SED. The standards for the infrared properties of local low metallicity galaxies are \citet{galliano2011} and \citet{remy2015}. The first reference shows theoretically how the temperature of the far infrared SED rises with increasing luminosity density. At very high luminosity densities, the mid infrared emission of small, stochastically heated grains becomes important and could broaden the overall SED if in an environment with a range of densities. In \citet{remy2015}, an increase in luminosity density fluctuations is shown to correlate empirically with increases in the maximum luminosity density, leading to a broadening of the far infrared SED with increasing luminosity, as expected. These works focus on nearby low luminosity galaxies over a range of metallicity, but their result resembles the behavior we find at high redshift and high luminosity. 

At high redshift,  \citet{burnham2021} model the increase in far infrared dust temperature in analogy with blackbody behavior\footnote{Although \citet{faisst2020} show that the emission is optically thin over much of the wavelength range of interest.}. They show that the increase correlates most closely with the surface density of star formation but only weakly with sSFR.  \citet{faisst2020} suggest that the effect is mostly confined to high redshift and is related to the decrease in metallicity and dust, so the emission is more optically thin and the hotter dust more readily observable. \citet{liang2019} suggest that the increase in temperature is strongly correlated with the sSFR, from theoretical considerations. \citet{mitsuhashi2024b} state that ``the observed redshift evolution of the dust temperature can be reproduced by an $\sim 0.6$ dex decrease in
the gas depletion timescale and $\sim 0.4$ dex decrease in the metallicity."

In comparison, we find that both the increase in temperature and the broadening of the SED at high redshift can be reproduced primarily based on the peak luminosity density  in the dusty regions. Since our models are optically thin, they implicitly include a large range of lower luminosity density with increasing distance from the heating source, which seems to be adequate to reproduce the behavior at the current level of constraints. Other influences such as metallicity have at most a secondary influence. 

Our study is a complement to that of \citet{sommovigo2025}, who developed a theoretical model for far-infrared SEDs that includes multi-temperature dust.  They concluded that, largely from ignoring the higher-temperature dust, single-temperature modified blackbody models could have errors in $L_{tot}$ up to $\sim$ 0.3 to 0.5 dex. The correction factors that we have derived range from 0.2 to 0.3 dex, derived from the shorter mid-infrared wavelengths that we capture empirically. They are largely in addition to those from their work,  since they result from the contributions of the much warmer dust and spectral features omitted from their calculations (see their Figure 1).

\section{Approach to Accurate IR Luminosities }
\label{sec:approach_to_accurate_IR_luminosities}


JWST is measuring SFRs and the sizes of many galaxies at high redshift, but the great majority are  below ALMA detection limits.  For example, where only photometry is available, SFRs can be estimated from $M_{\rm UV}$ \citep[e.g.,][]{calzetti2013}\footnote{ \url{https://ned.ipac.caltech.edu/level5/Sept12/Calzetti/Calzetti1_2.html}.}. Assuming no extinction, an approximate relation is SFR $\sim$ $3 \times 10^{-8} \times 10^{(-0.4 \times M_{\rm UV})}$, $\lesssim$ 1 M$_\odot$ yr$^{-1}$.  Even assuming this level of SF is fully embedded in dust, the continuum detection limits with ALMA cannot reach these galaxies  \citep{hill2024,liu2026}.  Instead, the typical masses, and hence potential SFRs, of galaxies detected in very deep ALMA observations are two to three orders of magnitude higher than the low-mass galaxies being studied in deep JWST surveys. 

The highest luminosity galaxies at $5 < z < 7$ are typically 1 kpc in diameter (FWHM) and if we assume the galaxies are spherical, their {\it average} luminosity density is typically $10^3 - 10^4$\,L$_\odot$\,pc$^{-3}$ \citep{spilker2016}.  These values are averages; presumably they  are non-spherical and also have unresolved internal structure, both of which would push the luminosity density in the most intensely far-infrared-emitting regions even higher. It is therefore no surprise that their far infrared SEDs differ significantly from those of typical lower redshift examples, where lower luminosity densities are prevalent. This accounts for their  average SED for $5 < z < 7$ resembling that of Haro 11 \citep{derossi2018}. 

The ongoing ALMA observations fall between these extremes, raising the question of how to estimate their infrared luminosities. We expect that the considerations discussed in this paper are important to obtain accurate estimates. The general increase in dust temperature has now been found in multiple studies, although in general they do not sample the width of the far infrared SED and thus cannot test its resemblance to that of Haro 11.  As discussed in the preceding section, there are a number of hypotheses for this behavior: \cite{ma2016,burnham2021,liang2019, mitsuhashi2024b,faisst2020}. For local galaxies,  the dependence of FIR dust SED on metallicity is very mild. Our models indicate that the dominant, by far, dependence is the luminosity density from the young newly formed massive stars (for local galaxies, see also \citealt{galliano2011} and \citealt{remy2015}). A further indication that metallicity plays little role at the redshifts of interest ($z$ $\sim$ 5 - 8) is that the high-redshift galaxies within range of ALMA do not have particularly low metallicity \citep{faisst2025,marszewski2025}. Our study therefore strongly favors the proposed dependence on star formation surface density, which is a proxy for high luminosity density.  

\begin{table*}
\label{morph_metrics}
\centering
\caption{A sample of high luminosity density galaxies} 
\begin{tabular}{|l|l|l|l|l|}
\hline Galaxy & $z$ & $L_{\rm IR}$ ($L_\odot$) & $r_{\rm eff}({\rm kpc})$ & Reference\\
\hline SPT-S J034640-5204.9 & 5.656 & $3.6 \pm 0.3 \times 10^{13}$ & $0.61 \pm 0.03$ & \citet{ma2016}\\
ADFS-27 & 5.655 & $2.4 \pm 0.3 \times {10^{13}}^a$ & $\sim 0.7, 0.7$  & \citet{riechers2017}\\
 ALMACAL-1 A1-A\tablenotemark{a}& 3.442 & $5 \times 10^{12}$ & $\sim 0.15$  & \citet{oteo2017}\\
 ALMACAL-2 A2-A\tablenotemark{a}& 3.442 & $3 \times 10^{12}$ & $\sim$ 0.15  & \citet{oteo2017}\\
  ASXDF1100.027.1 & 2.8$^{+0.48}_{-0.70}$ & $\sim 3 \times 10^{12}$ & $0.4^{+0.4}_{-0.3} $  & \citet{ikarashi2015}\\
  ASXDF1100.045.1 & $> 5.5$ & $\ge 4 \times 10^{12}$ & 0.3$^{+0.3}_{-0.2}$
  & \citet{ikarashi2015}\\
  ASXDF1100.090.1 & $>$ 3.2 & $\ge 2.5 \times 10^{12}$ & $0.4^{+0.4}_{-0.3}$ & \citet{ikarashi2015}\\
   ASXDF1100.110.1 &  4.98$^{+0.72}_{-3.14}$
 & $\sim 3 \times 10^{12}$ & 0.3$^{+0.4}_{-0.3}$ & \citet{ikarashi2015}\\
   ASXDF1100.230.1 &  3.50$^{+0.18}_{-0.40}$
 & $\sim 3 \times 10^{12}$ & $0.4^{+0.4}_{-0.3}$& \citet{ikarashi2015}\\
    P007+04 &  6.002
 & $4.5 \times 10^{12}$ & $\sim 0.3$& \citet{venemans2020}\\
      J109-3047 &  6.79
 & $1.3 \times 10^{12}$ & $\sim 0.35$& \citet{venemans2020}\\
P231-20 &  6.59
 & $10 \times 10^{12}$ & $\sim 0.25$& \citet{venemans2020}\\
 J2054-0005 &  6.04
 & $6.2 \times 10^{12}$ & $\sim 0.4$& \citet{venemans2020}\\
 J2318-3029 &  6.15
 & $6.3  \times 10^{12}$ & $\sim 0.35$& \citet{venemans2020}\\
 J2348-3054 &  6.90
 & $5.7  \times 10^{12}$ & $\sim 0.3$& \citet{venemans2020}\\
  MAMBO-9A &  5.85
 & $4  \times 10^{12}$ & $\sim 0.9$ & \citet{casey2019}\\
 AZTEC-1  &  4.342  & $1.9 \times 10^{13}$  &  1.1  &  \citet{tadaki2018} \\
\hline
\multicolumn{5}{l}

$^{a}${Two similar sources; We give the total luminosity and the radii roughly for each. }
\end{tabular}
\label{lumdenssumm}
\end{table*}

The increase in width of the high redshift, high luminosity density SED is demonstrated in the composite SED generated from far infrared measurements of 18 galaxies at $5 < z < 7$ in \citet{derossi2018, derossi2020}. This is  confirmed in our models of high redshift galaxies and is similar to the behavior modeled in detail for local galaxies by, e.g., \citet{galliano2011} and \citet{remy2015}. This, then, is also at some level the expected behavior of the more extreme infrared-bright galaxies at high redshift. 

Table~\ref{lumdenssumm} is a partial listing of  high-redshift galaxies known to have very high luminosity densities, in addition to the 18 that made up the composite Haro-11-like SED in \citet{derossi2018}. It emphasizes that the behavior is not confined to $z > 5$ but occurs for a range of redshifts.  Although they are a minority of all high-$z$ galaxies observed with ALMA, the incidence of this behavior is high enough that it should be taken into account when  converting ALMA measurements into estimates of star formation rates. However, the implications of the SED behavior with luminosity density are broader. For example, galaxies with apparently modest luminosity density at modest spatial resolution may break up into high density clumps at higher resolution with  high level luminosity densities \citep[e.g.,][]{chen2023, gimenez2023,alvarrez2023,delavega2025}. Such behavior might also explain the galaxies with hot-dust-dominated SEDs such as those reported by \citet{casey2009}.

\section{Estimating Obscured Star Formation}
\label{sec:estimating_obscured_star_formation}

\subsection{Different Approaches toward Estimating Obscured Star Formation}

The very early formation of dust and even of very dusty infrared-emitting galaxies shows that obscured star formation is significant in the early Universe \citep[e.g.,][]{gall2011,schneider2024}. This is confirmed by the pronounced peak in the far infrared in the cosmic infrared background \citep{dole2006}. Deep ALMA data are the most direct means to estimate obscured star formation  in individual galaxies. However, the ALMA data are typically available only in a few  bands,  sometimes far from the SED peak in the rest frame, and for limited (i.e., not necessarily unbiased) samples. This makes the approach to determining a fiducial SED challenging. Therefore, SED templates are used to estimate typical SFRs from infrared and submm photometry. Although individual galaxies may depart significantly from what seems to be the appropriate template,  the templates are designed to capture average behavior, not the range of behavior nor  detailed features of it. 

Purely theoretical template models of the infrared SEDs for $z$ $>$ 4 galaxies have poor fidelity \citep[e.g.,][]{jones2023}, or have to be based on minimal constraints and have huge predicted ranges of key output parameters such as dust temperatures \citep{sommovigo2022}, or are strongly dependent on simulations without observational confirmation at $z$ $>$ 4 \citep[e.g.,][]{liang2019}, all of which must be viewed with skepticism. 

Therefore, a more empirical approach is needed. One might use the SEDs of local LIRGs and ULIRGs to match the observations of high redshift galaxies of similar luminosity. 
This is not an ideal procedure because the local galaxies are the product of a different evolutionary sequence than the high redshift galaxies \citep[e.g.,][]{bromm2011,dayal2018}. Reaching the extremely high rates of star formation locally requires a merger that causes vast amounts of ISM gas and dust to settle into the core of the merger product. 
As a result, the mid-infrared can be substantially obscured, distorting the intrinsic SED in a way that does not seem to occur often at high redshift. 

 The SED templates of \citet{rieke2009} have already been tested  against high redshift galaxies \citep{rujo2013,lyu2016,derossi2018}. In the first study, it was shown that for 1 $<$ $z$ $<$ 2.8, the \citet{rieke2009} templates for log($L$)=11.25 - 11.75 gave the most accurate correspondence to total infrared luminosities from Spitzer and Herschel photometry from 24 - 500 $\mu$m.  \citet{derossi2018} carried out a more detailed comparison and found that the log($L$) = 11.25 template was strongly preferred over the log($L$) = 11.75 one for 2 $<$ $z$ $<$ 4, and that it was also preferred compared with templates from \citet{chary2001}, \citet{kirkpatrick2015}, \citet{schreiber2018}, and by a small margin over that from \citet{magdis2012}.  
 
 For 5 $<$ $z$ $<$ 7, the study of \citet{derossi2018} indicated that the average SED was much broader than the log($L$) = 11.25 template, and was better fitted by a Haro 11-like SED. The choice of a standard vs. Haro 11-like template could result in  substantial differences in estimating the total infrared luminosity and hence the star formation rates.  The following section addresses this critical choice. 

\subsection{Estimating $L_{\rm tot}$}

This paper has shown that the different behavior of the infrared SED is dominated by a single  key physical parameter, the luminosity density. With this insight, we review the general estimation of $L_{\rm tot}$, focusing on high redshifts, $z \ge 2$. 

If detailed far infrared photometry is unavailable, the common practice is to fit a modified blackbody to the observed-frame submm measurements  \citep[e.g.,][]{beelen2006, leipski2014}, i.e., a greybody of temperature $T$ with the emissivity proportional to $\nu^{\beta}$. In these cases, there is seldom sufficient information to fit templates in the rest mid-infrared and short-wavelength far infrared. \citet{casey2012} proposed adding a power law to fill in the shorter wavelengths across the mid-infrared. However, fitting this component involves three free parameters, i.e., requires at least three measurements at $\lambda \le$ 70 $\mu$m. The approach can still use  defaults for some of the parameters, but since at best there may be only a single measurement at the critical wavelengths (and none is more typical), the extra flexibility of this approach is not helpful in determining an accurate mid-infrared SED. 

Instead, we can illustrate the range of possibilities by adopting a modified blackbody, matching it to the far infrared peak of a template\footnote{Doing a full fit would be inadvisable because the templates are subject to internal errors and the modified blackbody is an abstraction, so the fit could be weighted toward the lower flux regions where the resulting uncertainties are large, whereas for luminosity we need to weight around the peak.} and determining the correction needed to obtain $L_{\rm tot}$ according to different assumptions. The correction is the ratio of the integration of a full template SED from 8 to 1000 $\mu$m to determine the luminosity associated with it, divided by  the result of a similar integration of the modified blackbody fitted to the template{\footnote {Experiments showed that the results depend on exactly what wavelength range is selected for the fitting and luminosity determination. To be specific, our results are based on a least-squares fit to the template from 70 to 1000 $\mu$m with variables temperature and $\beta$, compared with a modified blackbody over the same wavelength range. The SED of Haro 11 is too broad for a single blackbody to provide a good fit.  }. Some examples can be found in Table~\ref{tab:modBBcorr}.  These cases are meant as examples to illustrate the size of the ``missing'' flux with single blackbodies and how it varies with template. The errors in determining the parameters of the blackbody fits on real data are generally much larger than these corrections.

\begin{figure}[h!]
\begin{center}
\vspace{-0cm}
        \hspace{-0cm} \includegraphics[width=8.5cm]{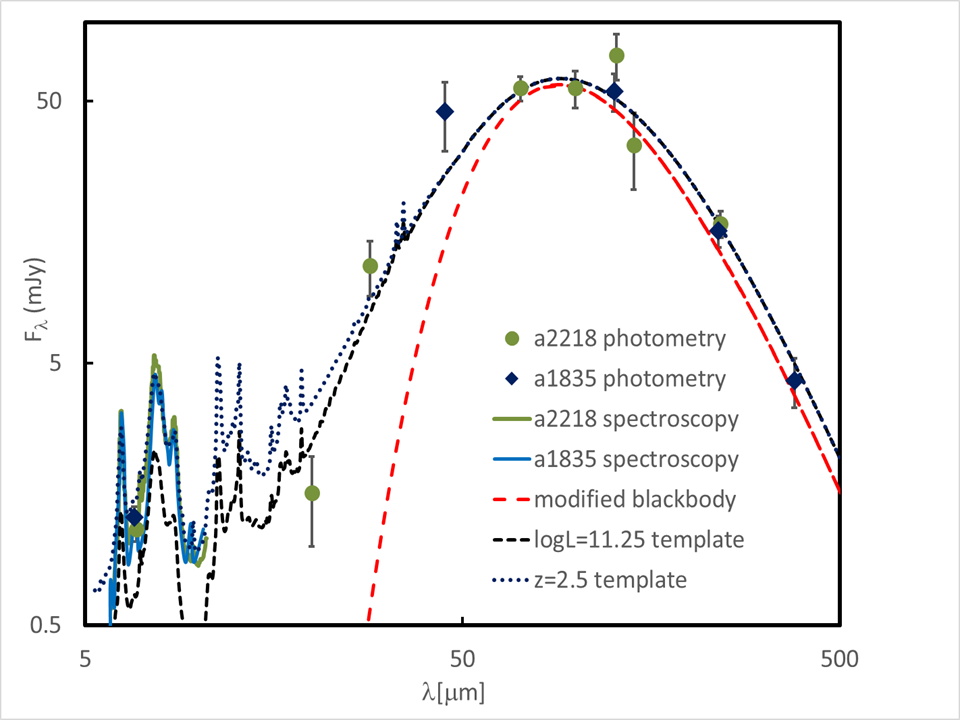}
        \hspace{-0cm} 
\end{center}
\caption{Photometry and spectroscopy of A1385a and A2128a merged, together with a customized template fit (the ``z = 2.5 template''). See text for details.}
\label{fig:template25}
\end{figure}

As a check on the use of the log($L$) = 11.25 template, we will construct a reference SED and determine a correction from modified blackbodies to total luminosity for two {\it individual} galaxies at $z \sim 2.5$, A1835a and A2218a \citep{rigby2008}. Both are lensed, providing access to lower luminosity than in other samples. A2218a is at $z$ = 2.516 and has intrinsic $L_{\rm tot} \sim 8 \times 10^{11}$ L$_\odot$, with a magnification of $\sim$ 22 \citep{rigby2008}. For A1835a, the intrinsic $L_{\rm tot} \sim 4 \times 10^{12}$ L$_\odot$  at $z$ = 2.56 with a magnification\footnote{We have refined the luminosity comparison through our comparison of the two SEDs.} of 3.5. Thus, a composite from the two corresponds to $L_{\rm tot}$ slightly above 10$^{12}$, significantly lower than the luminosities of the $z$ $\sim 2-3$ galaxies used in previous comparisons \citep{rujo2013, derossi2018}. Measurements of these galaxies are listed in Table~\ref{fluxes}, and plotted in Figure~\ref{fig:template25}, after multiplying those for A1835a by a factor of 1.23 (determined from the fit to the template).   In addition to the traditional photometric points each galaxy has a low resolution spectrum with about 50 spectral points from 6 to 11 $\mu$m. As shown in the figure, they indicate similar SEDs, and in the following we treat them together. 

Although the $\log(L) = 11.25$ template is a good fit in the far infrared {\bf (i.e. $\lambda > 50~\mu{\rm m}$)}, it falls below the measurements in the mid-infrared  (6 - 11 $\mu$m).  This behavior is similar to the trend with luminosity seen locally \citep{rieke2009}. We modify the template by fitting an interpolated one for $\log(L) = 10.85$ short of 50 $\mu$m, as shown in the figure (dotted blue line).  To do so, we normalize the SED in the 100 - 400 $\mu$m range and then modify it by replacing the values short of 60 $\mu$m with a lower luminosity template, requiring that it join smoothly to the log(L) = 11.25 one and that $\chi^2$ in the 6 - 11 $\mu$m range has been minimized.  The  correction to $L_{\rm tot}$  with this template is in Table ~\ref{tab:modBBcorr}. It generally agrees with the similar parameters for the $\log(L)=11.25$ template, supporting its applicability at $z\sim2.5$.

These two cases suggest a nominal temperature of $\sim 35$ K for this redshift range. This can be compared with the summary in \citet{mitsuhashi2024b}, Figure 9. The model from \citet{sommovigo2022}
predicts a temperature of 34 K at $z$ = 2.5, while the model of \citet{liang2019} indicates 35 K. These agree well with the temperatures of the two relevant templates. 

For the redshift range 3 $<$ $z$ $<$ 4, the predicted far infrared temperatures at $z$ = 3.5 are 37.5 K \citep{sommovigo2022} and 40 K \citep{liang2019}. This suggests that the $\log(L) = 11.5$ template  might be a good match as shown in Table~\ref{tab:modBBcorr}. The reduced correction factor for this template is, however, largely from the high optical depths in the mid-infrared, which do not occur for the high redshift galaxies \citep[e.g.,][]{shipley2016,florian2025}. We instead modify the original template for application at high redshift as we did for A1385a and A2218a, i.e., joining the interpolated template for $\log(L) = 10.85$  to the $\log(L) = 11.5$ template at 50 $\mu$m. The indicated correction factor is 1.7. That is, the necessary correction does not differ substantially from the $\sim$ 1.6 indicated for the 2 $<$ $z$ $<$ 3 range.

At $z$ = 5.5, the indicated nominal dust temperatures are 43\,K \citep{sommovigo2022}, or 49 K (extrapolation of \citealt{liang2019}). The parameters in \citet{venemans2020} seem appropriate for nominal fits, i.e., $T$ = 47 K and $\beta$ = 1.6. For $z$ $>$ 4, high luminosity densities become more common, see Table~\ref{lumdenssumm}. To evaluate whether they yield a shift in the relation to $L_{\rm tot}$, we refer to \citet{mitsuhashi2024b}, figure 5. It shows that for luminosities $\ge$ $10^{12}$ L$_\odot$, and particularly for those approaching or exceeding $10^{13}$ L$_\odot$, the luminosity densities are often, perhaps usually, in the Haro~11 class and the correction from a modified blackbody fit to $L_{\rm tot}$ should be increased to a factor of up to $\sim$ 2. However, for lower luminosities, typical luminosity surface densities are an order of magnitude lower \citep{mitsuhashi2024b}, and the correction factor will resemble those for lower redshifts.

\begin{table*}
\label{morph_metrics}
\centering
\caption{Fluxes (mJy) for A2218a and A1835a} 
\begin{tabular}{|l|l|l|l|l|}
\hline band & A2218a & ref & A1835a & ref\\
\hline 
PAH spectrum & -- &  \citet{rigby2008}  &  ---  & \citet{rigby2008} \\
24 $\mu$m &  1.16 $\pm$ 0.12 &
\citet{rigby2008}  &  0.99 $\pm$ 0.10  &  \citet{rigby2008} \\
70 $\mu$m &  1.6 $\pm$ 0.6  & \citet{papovich2009}\tablenotemark{a} &  &  \\
100 $\mu$m & 12 $\pm$ 3 &  \citet{marton2024}  &  & \\
160 $\mu$m &  &   & 35 $\pm$ 10 & \citet{marton2024} \\
250 $\mu$m & 55.5 $\pm$ 6.1 &  \citet{finkelstein2011} &  &  \\
350 $\mu$m  & 56 $\pm$ 9.4 & \citet{finkelstein2011} &  &  \\
450 $\mu$m  & 75 $\pm$ 15 & \citet{kneib2004} & 41. $\pm $ 6.9 & \citet{ivison2000} \\
500 $\mu$m  & 34 $\pm$ 10.5 & \citet{finkelstein2011} &  &  \\
850 $\mu$m & 17 $\pm$ 2 & \citet{kneib2004} & 14.6 $\pm$ 1.8  & \citet{ivison2000} \\
1350 $\mu$m  &  &  & 3.3 $\pm$ 0.7  &  \tablenotemark{b} \\
\hline
\multicolumn{5}{l}
{$^a$ corrected to remove contribution of component A}
{$^b$ average of values from \citet{ivison2000, downes2003} }
\end{tabular}
\label{fluxes}
\end{table*}

\begin{deluxetable*}{lcccc}
\tabletypesize{\footnotesize}
\label{sfrlist}
\tablecaption{Corrections from FIR Luminosity to total Infrared Luminosity} 
\tablewidth{0pt}
\setlength{\tabcolsep}{1.5pt}
\tablehead{
\colhead {template} & 
\colhead  {T(K)}  &
\colhead{$\beta$}&
\colhead {correction}  &
\colhead {reference}  \\
}
 
\startdata
log(L)=11.25\tablenotemark{a}  &  34.8 &  1.51  &  1.63 &  \citet{rieke2009}\\
a2218a 
+ a1835 &  35.0  &  1.50  &  1.74  &  see text \\
log(L)=11.5  &  40.9  &  1.35  &  1.39  &  \citet{rieke2009}\\
z = 6  & 47 & 1.6  &  1.8  & \citet{schreiber2018}   \\
Haro 11  &  47  &  1.6  &  $\sim$ 2  & \citet{derossi2018} \\
\hline
\enddata
\tablenotetext{a}{The nominal template for z $\sim$ 2 - 2.8 \citep{rujo2013, derossi2018}}
\label{tab:modBBcorr}
\end{deluxetable*}

\section{Conclusions}
\label{sec:conclusions}

Contributing to astronomy's long-standing pursuit to map the totality of cosmic star formation, we address a key physical effect in assessing the contribution of dust-obscured star formation, the higher-temperature and broader (in wavelength range) far-infrared SEDs found for high redshift ($z \sim 6$) star forming galaxies. Specifically, we demonstrate the following points: 

\begin{itemize}

\item{Such SEDs occur locally and are a product of very high luminosity density in the ISM of a galaxy, with relatively little dependence on parameters such as metallicity.} 

\item{Reducing the luminosity density results in SEDs more typical of those observed for ``normal'' local infrared galaxies, i.e., ones that are at lower temperature and narrowed in wavelength range.}

\item{Models addressing the behavior of very high redshift galaxies demonstrate this same behavior.}

\item{Many of the {\it most luminous} high redshift galaxies accessible with ALMA {\bf are expected to} have warm dust components reflected in broad SEDs.}

\item{However, this behavior is confined to the most luminous examples; recent ALMA detections extend down to luminosities where the broadening should not be expected.}

\item{We have analyzed how to correct modified blackbody values to total infrared luminosities for typical galaxies. For infrared galaxies at $z$ $>$ 2, luminosities estimated with modified  blackbodies and standard parameter choices need to be multiplied by a correction factor of $\sim$ 1.6 - 1.7 to convert them to total infrared luminosity.}

\item{For galaxies at luminosities $\ge$ $10^{12}$ L$_\odot$,  a correction factor up to $\sim$ 2 (for the highest luminosities) is more appropriate. }

\end{itemize}

In terms of \ broader context, our results contribute to the challenge of harnessing JWST and ALMA  observations at the high-redshift frontier to ``stress-test'' the $\Lambda$CDM cosmological model \citep[e.g.,][]{liu2022,boylan2023}. This crucial test relies in particular on the most massive, star forming systems at cosmic dawn, where dust obscuration is a key uncertainty \citep[e.g.][]{ferrara2025}. For a proper accounting of the high-mass end of the galaxy luminosity function, we provide the necessary improved understanding of the (rest-frame) far-infrared SED. 

\section*{Acknowledgements}

  Work on this paper was supported in part by grant 80NSSC18K0555, from NASA Goddard Space Flight Center to the University of Arizona. 
We thank Alexander Ji for providing tabulated dust opacities for some of the
dust models used here. This work makes use of the Yggdrasil code \citep{zackrisson2011}, which adopts Starburst99 SSP models, based on Padova-AGB tracks \citep{leitherer1999, vazquez2005} for Pop~II stars. The work of C.C.W. is supported by NOIRLab, which is managed by the Association of Universities for Research in Astronomy (AURA) under a cooperative agreement with the National Science Foundation.

\bibliographystyle{aasjournal}

\end{document}